\documentclass[aps,preprint]{revtex4-1}
\usepackage{graphicx}
\usepackage{amssymb}
\usepackage{amsmath}
\usepackage{color}
\usepackage{enumerate}
\usepackage{placeins}
\usepackage{bm}
\usepackage{floatrow}
\usepackage{subcaption}
\begin{document}

\title{Twitching Motility of Bacteria with Type IV Pili: \\Fractal Walks, First passage time 
and their Consequences on Microcolonies}
\author{Konark Bisht$^1$, Stefan Klumpp$^2$, Varsha Banerjee$^1$, Rahul Marathe$^1$}
\email{maratherahul@physics.iitd.ac.in} 
\affiliation{$^1$Department of Physics, Indian Institute of Technology, Delhi, Hauz Khas 110016, New Delhi, India.\\
$^2$ Institute for Nonlinear Dynamics, Georg-August University G\"ottingen, Friedrich-Hund-Platz 1, 
37077 G\"ottingen, Germany.}

\begin{abstract}
A human pathogen, \textit{Neisseria gonorrhoeae} (NG), moves on surfaces by attaching and retracting polymeric structures called Type IV pili. The \textit{tug-of-war} between the pili results in a two-dimensional stochastic motion called \textit{twitching motility}. In this paper, with the help of real time NG trajectories, we develop coarse-grained models for their description. The \textit{fractal properties} of these trajectories are determined and their influence on \textit{first passage time}  and formation of bacterial  microcolonies is studied. Our main observations are as follows: (i)  NG performs a fast ballistic walk on small time scales and a slow diffusive walk over long time scales with a long crossover region; (ii) There exists a characteristic persistent length $l_p^*$ which yields the fastest growth of bacterial aggregates or biofilms. Our simulations reveal that $l_{p}^{*} \sim L^{0.6}$, where $L\times L$ is the surface on which the bacteria move; (iii) The morphologies have distinct fractal characteristics as a consequence of the ballistic and diffusive motion of the constituting bacteria.
\end{abstract}

\maketitle

\section{Introduction} Bacteria crawl, glide and twitch over surfaces with the help of {\it pili} \cite{JKF,JMS,CLG,JH,JSM}. These are micrometer sized hairy appendages, composed of oligomeric pilin proteins. Particularly in two dimensions bacterial motility is induced through cycles of polymerization and depolymerization of type IV pili (T4P). While polymerization causes extension of the pilus and subsequent adhesion to the surface, depolymerization induces retraction, release and propulsion. Often, several pili act cooperatively \cite{NB,RM,DLH}. The motion of the bacterium in this case, as mimicked by the 
stochastic {\it tug-of-war model}, is determined by a vectorial balance of forces between the retracting pili \cite{RM,CH,DO,VZ}. Typically, a single pilus can generate a force up to $100$ pN while a bundle of pili up to $1 - 2$ nN and the velocities for surface motility fall in the range $1 - 2$ $\mu$ms$^{-1}$ \cite{NB,AJM,MJM}. As the number of pili involved in the tug-of-war mechanism is variable, there is an element of stochasticity in the direction of motion and distances traversed by a bacterium in a polymerization-depolymerization cycle. Often these distances are significantly longer than the length of a single pilus. The resulting irregular and jerky movement is usually referred to as {\it twitching motility} \cite{HL}. Experiments reveal a correlation between multiple pili coordination and directional persistence. The {\it persistence time}, over which the direction of motion remains unchanged, exhibits a pronounced rise with increasing number of pili \cite{CH}. Some prominent examples of bacterial species with T4P are \textit{Pseudomonas aeruginosa} responsible for fatal airway infections, \textit{Neisseria gonorrhoeae} leading gonorrhoea, \textit{Escherichia coli}, \textit{Salmonella} and \textit{Shigella} inciting bladder and intestinal infections, \textit{Neisseria meningitides} causing the deadly meningitis and the soil predator \textit{Myxococcus xanthus}. A recent study on \textit{P. aerugonisa} shows that  the number of T4P and their angular distribution on bacterial surface determine their motility phases \cite{YBK}.

Persistence and change of direction in the walk allow the bacteria to explore the surface in search of food, assembling into microcolonies, formation of biofilms, guided tissue invasion and other pathogenesis-related events. It is logical to expect that these search processes are strongly influenced by the properties of the random walk. An important quantifier of a random walk can be obtained by evaluating the so-called Brownian trace \cite{PA}. It yields the {\it Hurst exponent} $\alpha$, $0 < \alpha \le 1$, which is a measure of the roughness of the trajectory and has important implications on diffusion. For example, $\alpha = 1/2$ implies regular diffusion while $\alpha = 1$ is a sign of ballistic diffusion. The motion is superdiffusive and persistent  if $1/2<\alpha<1$, while sub-diffusive and anti-persistent if $0<\alpha<1/2$. The Hurst exponent also yields the fractal dimension: $d_f = d-\alpha$, where $d$ is the Euclidean space dimension.  There are many questions of relevance in the context of bacterial walks. (a) What is the distribution of step lengths, Hurst exponent and the fractal dimension? (b) Do these characteristics change as the walk becomes longer and longer? (c) How do they affect the First Passage Time (FPT)? The FPT \cite{SR} is the time taken by the bacterium to reach a target such as food, a bacterial colony or a weak spot of the host tissue and could be important for the speed of biofilm formation and the spread of infections \cite{MRE1,SNM,MRE2,LK,SD}.
(d) Is there a correlation between the fractal dimension of the bacterial walk and morphological features of the colonies and biofilms? Our work addresses some of these questions.

In this paper, we use experimental trajectories traced by \textit{Neisseria gonorrhoeae} (NG) which move on a glass substrate. Each trajectory is of few minutes duration with temporal resolution of $0.1$ s.
On these time scales, the bacteria are still away from the substrate edge, and the tracks resemble random walks. We analyze them to obtain Hurst exponent and corresponding fractal dimensions. This information is then used to construct models, which mimic motility of twitching bacteria e.g. NG.  Monte Carlo (MC) methods are used to simulate bacterial walks on an $L \times L$ surface to  study the influence of the walk characteristics on the FPT and the formation of colonies and biofilms. Their morphological characteristics such as the fractal dimensions are evaluated. The main results of our study are as follows: \\
(a) The step lengths in the NG trajectories obey an exponential distribution: $P(l) = (1/l_p)~\exp\left(-l/l_p\right)$ with the persistence length $l_p\simeq 2$ $\mu$m \cite{CH}.\\
(b) The Hurst exponent evaluated from these tracks exhibits a long crossover from a ballistic regime ($\alpha \simeq 1.0$) to a diffusive regime $(\alpha \simeq 0.5)$, as the window of observation increases. \\ 
(c) The walks are {\it fractal} and {\it persistent}. The bacterial motion is a superposition of fast ballistic diffusion (directed motion) on small time scales and slow diffusion (random walk) over longer time scales. \\
(d) Computer simulations reveal that for each value of $L$, there is an optimal value of the persistence length $l_p^*$ for which the first passage time (FPT) is the least. \\
(e) The bacterial colony resembles a diffusion-limited-aggregation (DLA) cluster in the early stages. As the cluster grows, the morphology at the boundary loses its delicate feathery character. The core and the periphery are characterized by different fractal dimensions. These observations are a consequence of the observed crossover in the Hurst exponent.\\
(f) While the DLA backbone results from the long diffusive walks ($\alpha \simeq 0.5$), the smearing is due to ballistic walks ($\alpha \simeq 1.0$) which are executed on shorter time scales.\\

A large volume of work has emerged to understand the role played by the local environment on the formation of bacterial microcolonies and biofilms \cite{LK,GAO}. One of the main reasons for this intensive research activity is because they offer protection to pathogens from antibiotic attacks, and are at the root of chronic infections \cite{JWC,LHS}. In this context, understanding the formation and morphology of biofilms could help optimize strategies for their removal. To the best of our knowledge, there are no experimental or theoretical studies which correlate the Hurst exponent and fractal dimension with FPT and the subsequent formation of aggregates and biofilms. Our analysis therefore provides fresh insights on this topic using simple methodologies from statistical physics. We emphasize that they are generic to the wide class of micro-organisms with T4P exhibiting twitching motility.

The rest of the paper is organized as follows. In Section 2,  we provide experimental trajectories showing the characteristic twitching motility of NG. Evaluations of step-length distribution, persistence length, Hurst exponent, first passage time (FPT), etc. are also provided in this section. Section 3 discusses our model and boundary conditions which mimic the twitching motility of prototypical NG. Computer simulations to understand the role played by the persistence length and Hurst exponent on FPT,  growth of bacterial colonies and their morphologies are also discussed. We also compare our inferences from simulations with a variety of bacterial colonies observed in laboratory experiments. Finally in Section 4 we provide a summary and discussion of our results.

\section{Analysis of recorded NG trails}

\begin{figure}[!htbp]
\begin{subfigure}{0.5\linewidth}
\centering
 \includegraphics[scale = 0.25]{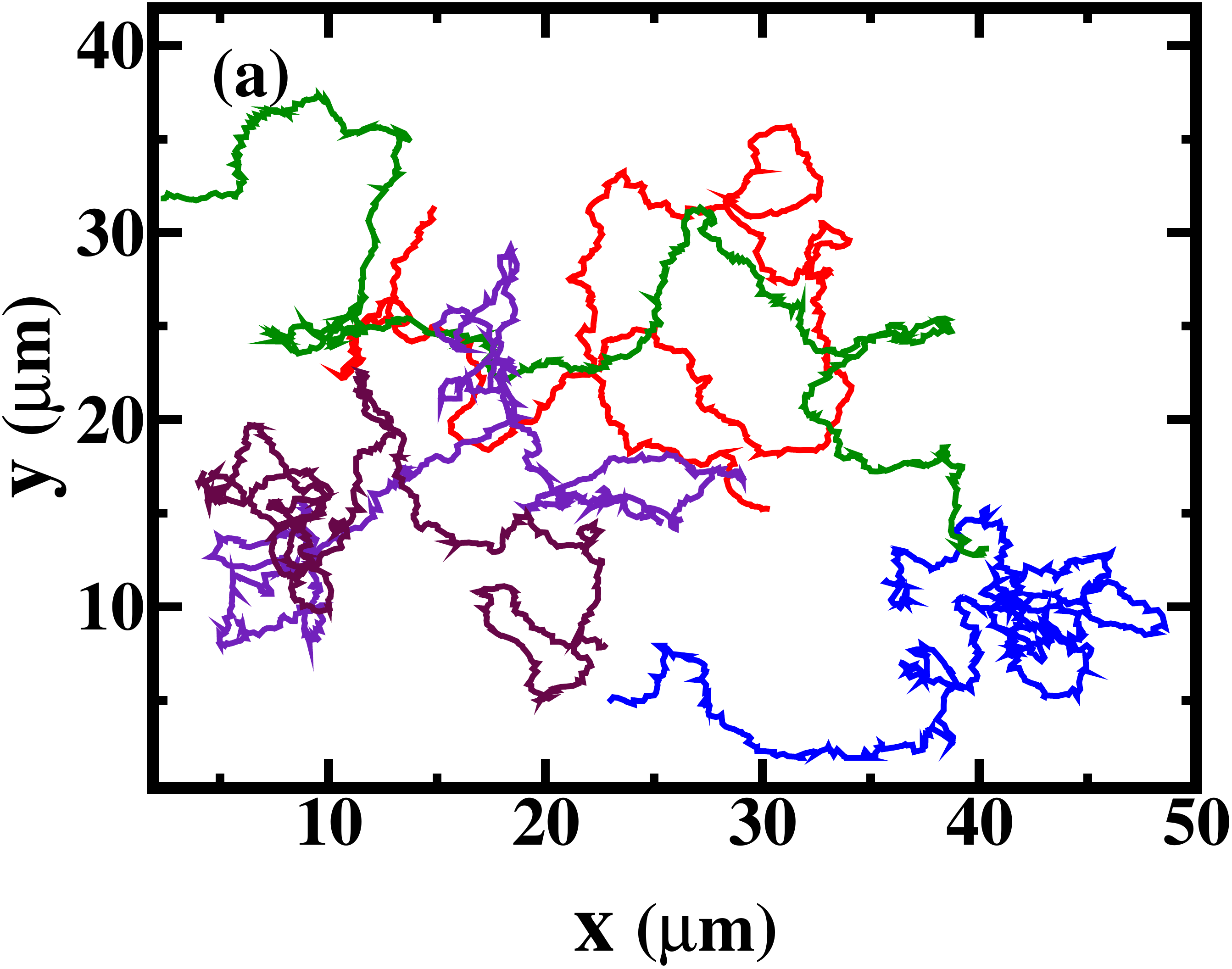}
\end{subfigure}
\begin{subfigure}{0.49\linewidth}
\centering
 \includegraphics[scale = 0.25]{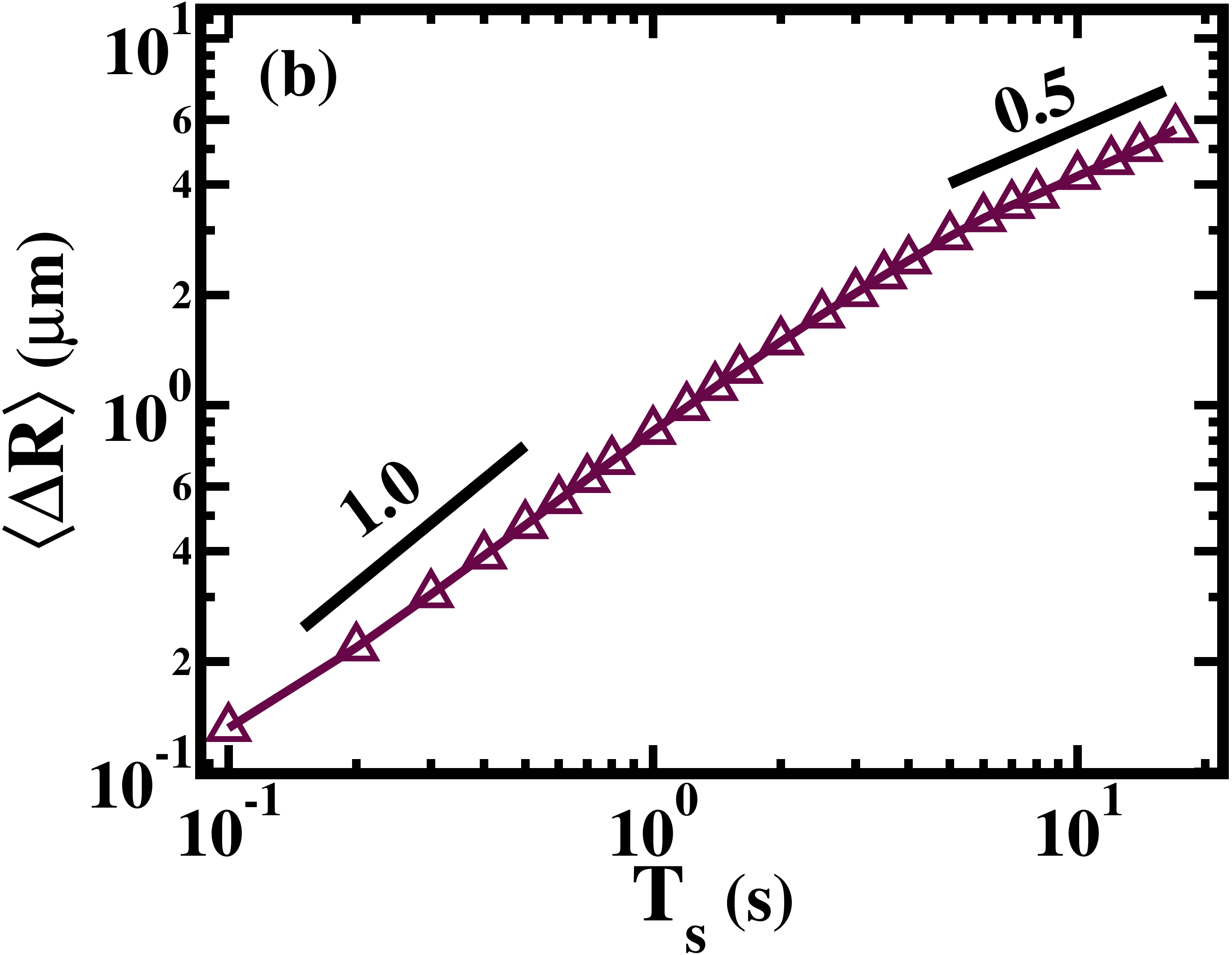}
\end{subfigure}
 \caption{(a) Trajectories of a \textit{Neisseria gonorrhoeae} (NG) bacterium on a BSA-coated glass surface. (b) Brownian trace $\langle \Delta R\rangle$ vs. $T_s$ for the experimental trajectories as discussed in the text. Solid lines indicate the value of the Hurst exponent $\alpha$.}
 \label{fig:expTrack}
\end{figure}

We first analyzed experimental trajectories of NG on glass slides from the experiments of B. Maier's group \cite{CH, NK, BM}. The trajectories were kindly provided by C. Meel and B. Maier. In these experiments, movies tracking their positions were recorded by a microscope with a temporal resolution of $0.1$ seconds for $2$ minutes. The details of sample preparation and the tracking experiment can  be found in reference \cite{CH, NK, BM}. A few of the NG motility tracks are provided in Fig.~\ref{fig:expTrack}(a). Notice that they are directed on a local scale, but generally random and irregular. Reports indicate that the bacterium moves with the velocity $v \simeq 1.5$ $\mu$ms$^{-1}$ \cite{DO}. 

We adopt the following procedure to calculate the step length distribution $P(l)$ vs. $l$ from the bacterial tracks. A new step is assumed if the directional change is $\gtrsim45^\circ$ \cite{CH}. The step length $l$ is then the distance travelled by the bacterium before this change. We find that the distribution is:
\begin{equation}
\label{exp}
P(l) = \frac{1}{l_p}\ \exp\left(-l/l_p\right),
\end{equation}
consistent with the observations made in reference \cite{CH} from an analysis of NG tracks.  The parameter $l_p$ is referred to as the persistence length. Note that for this exponential distribution, the mean and the standard deviation are equal to $l_p$. The latter has a strong dependence on the number of pili on the bacterial cells. In wild type NG bacteria, the average number of pili is $\simeq 7$  \cite{CH,RM}. Consequently, $l_p$ is significantly longer than the average pilus length of $1~\mu$m. The average number of pili for {\it E. coli}, {\it C. xerosis} and {\it C. pyogenes} is $\simeq 10$, while that for {\it C. kutscheri}, {\it C. diphtheriae} and {\it C. pseudodiphtheriticum} is $\simeq 100$ \cite{CH, RY}. We expect $l_p$ to be much larger for these species.

Next we quantify the irregularity in the bacterial tracks. A convenient measure of roughness is the Hurst exponent. The latter is evaluated by constructing the {\it Brownian trace.} Suppose the position of the random walker at time $t=0$ is the origin: $\left(x(0),y(0))=(0,0\right)$. If $R(t)$ is the distance of the walker at instant $t$, then $R(t) =  \left(x(t)^2+y(t)^2\right)^{1/2}$. Consider the set of {\it all} pairs $(t_i,t_j)$ separated by $T_s$, i.e. $\mid t_{i} - t_{j}\mid = T_s$ $\forall \ i,j$. The Brownian trace is then defined as:
\begin{equation}
\label{trace}
\langle \Delta R(T_s)\rangle = \langle \mid R(t+T_s) - R(t) \mid \rangle,
\end{equation}
where the angular brackets indicate an average over the {\it all} the time intervals of length $T_s$. 
Further, the trace generally obeys the form
\begin{equation}
\label{trace1}
\langle \Delta R(T_s)\rangle  \propto T_s^{\alpha},
\end{equation}
where $\alpha$ is the Hurst exponent. It conveys important properties of the Brownian trajectory. $\alpha>0.5$ indicates persistence while $\alpha<0.5$ indicates anti-persistence. While the former is the tendency of the walker to maintain the direction of motion in subsequent steps, the later implies directional reversal in subsequent steps. Further, a slow, diffusive (regular) random walks has a Hurst exponent $\alpha = 0.5$. Ballistic walks of super-diffusion is characterized by $\alpha > 0.5$, while sub-diffusion by $\alpha<0.5$. 

In Fig.~\ref{fig:expTrack}(b) we plot $\langle \Delta R(T_s)\rangle$ vs. $T_s$ averaged over $10$ NG trajectories on a log-log scale. For small values of  $T_s \sim 1$ s, the Hurst exponent $\alpha \simeq 1$. On short time scales, the walks are persistent, super-diffusive and fractal. At larger values of $T_s$, there is a crossover to $\alpha \simeq 0.5$ indicating that the coarse-grained motion is akin to a regular diffusion. These observations suggest that the bacterial walk is a superposition of a faster {\it directed} motion on small time scales and a slow {\it regular diffusion} or random walk over longer time scales. We note that in a recent study on the random walks of {\it P. Aeruginosa} \cite{YBK} has also characterized different regimes of motility by different 
exponents. Compared to two motility regimes that we see here, additional regimes corresponding to 
self-trapped localization and sling shot motion are also observed in {\it P. Aeruginosa} which are not seen in NG.

\section{Model and Simulation Results}
\subsection{First Passage Time:}
The features of experimental trajectories provide us with guidelines to construct a coarse grained model to mimic bacterial motility. Our model does not take into account the details of tug-of-war mechanism. Rather, it relies on walk characteristics such as the step length distribution, persistent time and Hurst exponents. The bacterium, moving with the velocity $v = 1.5 \ \mu$ms$^{-1}$, executes a random walk confined to a $L\times L$ surface. The simple confinement that we consider here can be generated experimentally \cite{DO}. In addition, bacteria may face such confinements in their natural environment either due to walls, for example in porous materials, or due to surfaces with different chemistries. The step lengths $l$ are drawn from the exponential distribution of Eq.~(\ref{exp}) and the direction of motion is selected randomly from the uniform distribution $[0,2\pi)$. The time taken for a step of length $l$ is therefore $l/v$. In our model the motion is subjected to the condition that the movement along the boundary continues with the parallel component of the velocity, from the first hit until a direction leading to the interior is selected in the course of the random walk. This is one of the standard boundary condition for active Brownian particles \cite{PR}. We have performed our simulations for several values of $L$ ranging from $30$ to $2000$ $\mu$m. We choose a large variation for $l_p$ from $0.1$ to $100$ $\mu$m to understand the properties of random walks with exponentially distributed step lengths in general and bacterial walks in particular. All data unless mentioned has been averaged over $10^{6}$ walks. 

\begin{figure}[!htbp]
\vspace{0.5cm}
 \includegraphics[scale=0.3]{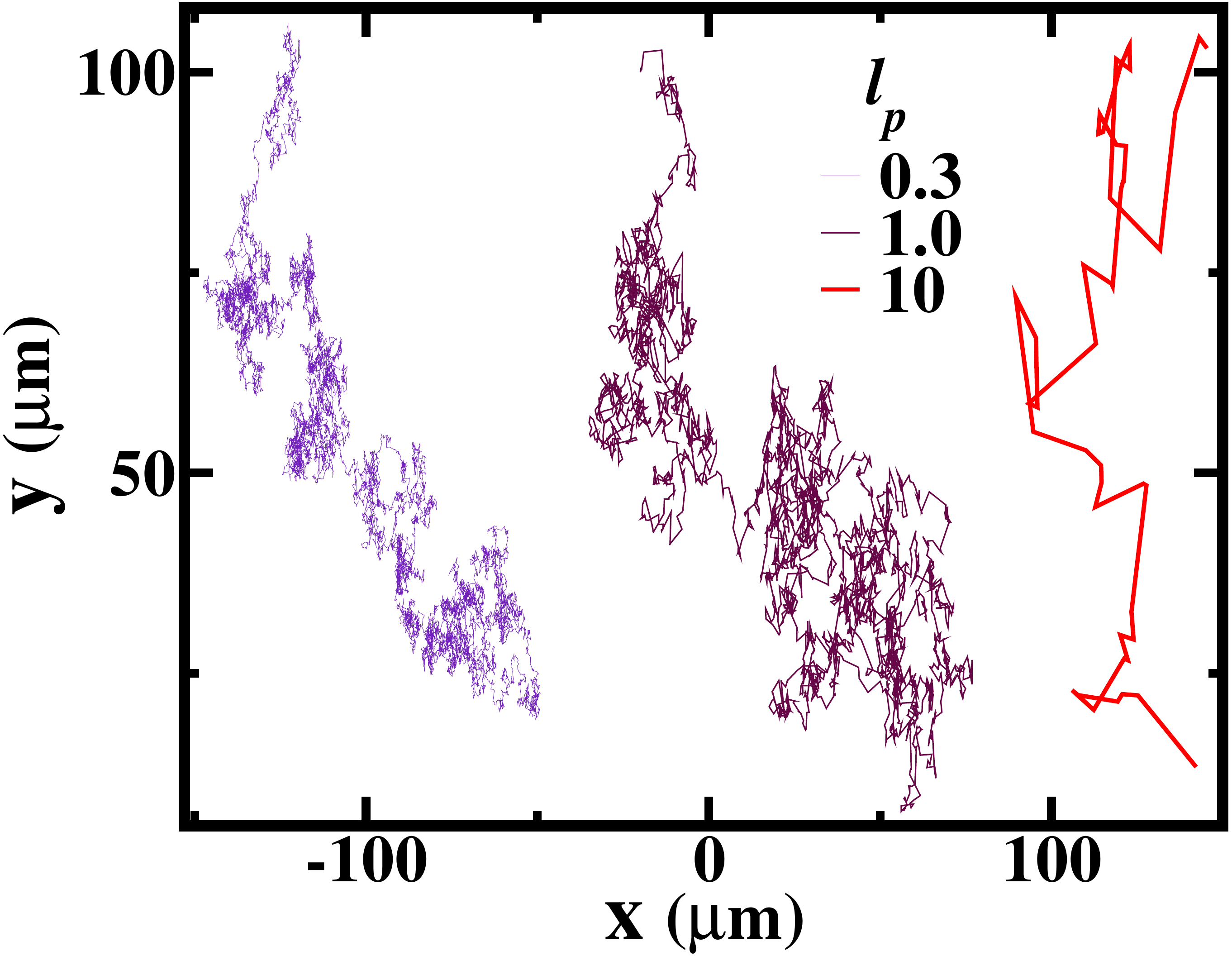}
 \caption{Trajectories from simulations for persistence length $l_p$ $(\mu$m) $= 0.3$, $1$ and $10$ on a $L\times L$ surface with $L= 200$ $\mu$m. The movement is well within the surface for observation timescales which run over 5 minutes.} 
  \label{fig:simTrack}
\end{figure}

\begin{figure}[!htbp]
\begin{subfigure}{0.34\linewidth}
\centering
 \includegraphics[scale = 0.2]{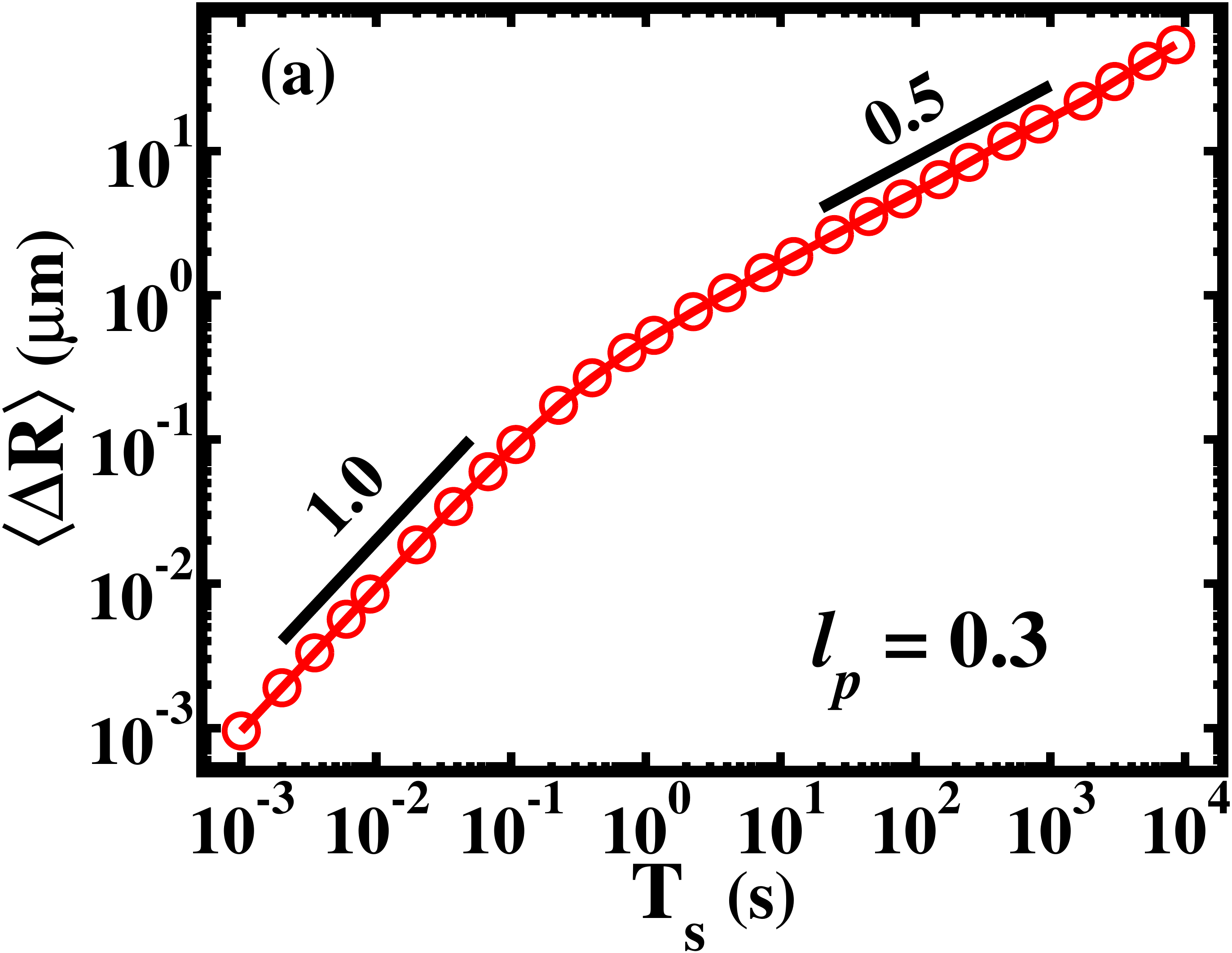}
\end{subfigure}
\begin{subfigure}{0.32\linewidth}
\centering
\includegraphics[scale = 0.2]{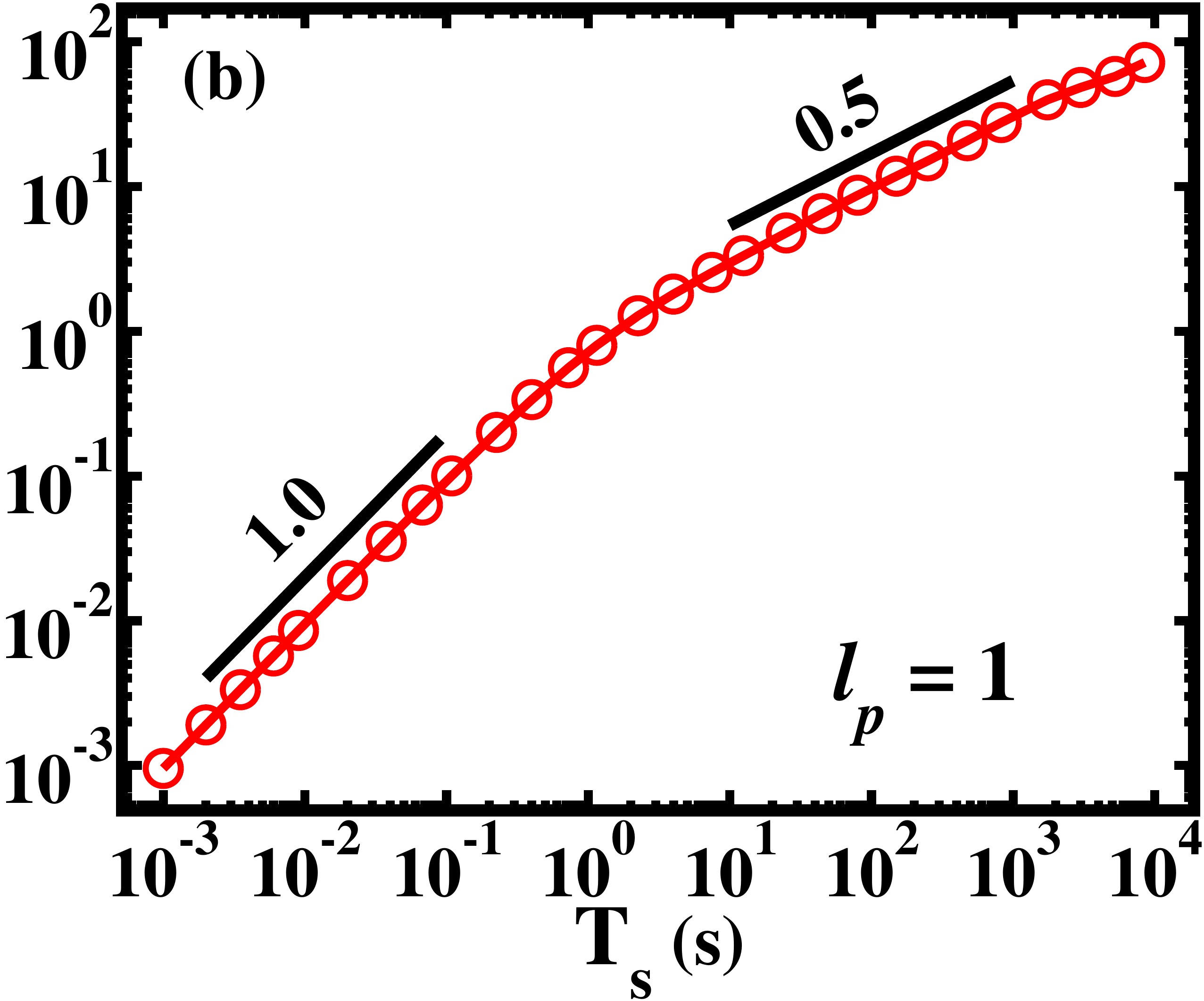}
\end{subfigure}
\begin{subfigure}{0.32\linewidth}
\centering
\includegraphics[scale = 0.2]{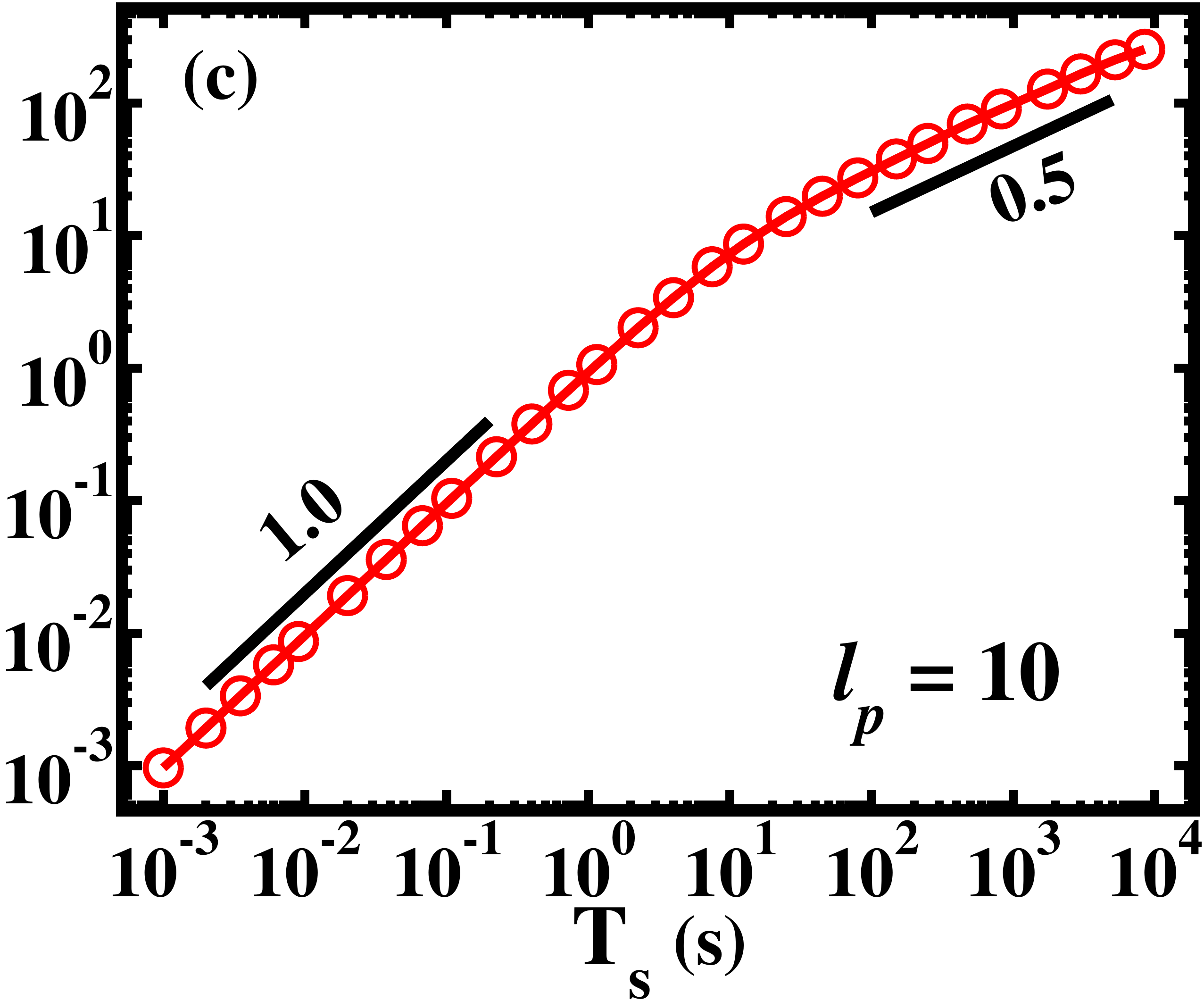}
\end{subfigure}
\caption{Plot of $\langle \Delta R\rangle$ vs. $T_s$ for trajectories from simulations for $L = $ 200 $\mu$m for (a) $l_p = 0.3$ $\mu$m, (b) $l_p = 1$ $\mu$m and (c) $l_p=10$ $\mu$m. The data mimics the cross-over from ballistic motion ($\alpha \simeq 1$) at smaller times ($\sim10^{-1}$ s) to diffusive motion ($\alpha \simeq 0.5$) at longer times  ($\sim10$ s) as observed experimentally. This crossover occurs at $T_s\sim l_p/v$.}
\label{fig:hurstS}
\end{figure}

Using the above model, we now study the connection between persistence length $l_p$, Hurst exponents and system size on the FPT. As discussed earlier, the latter is very consequential for search of food, spread of diseases and formation of microcolonies. It is an important quantity in the context of target search and provides a timescale for spread of infection or formation of microcolonies. 
Fig.~\ref{fig:simTrack} shows prototypical trajectories traced by the particle for $l_p$ = 0.3, 1 and 10 for $L$ = 200 $\mu$m. Over the observation timescale, we find that the bacterium is well within the $L\times L$ region, thereby mimicking the experimental walks. A plot of $\langle \Delta R(T_s)\rangle$ vs. $T_s$ on a log-log scale is depicted in Fig.~\ref{fig:hurstS}. The solid lines have slopes 1.0 (small $T_s$) and 0.5 (large $T_s$). All three data sets yield a Hurst exponent $\alpha \simeq 1.0$ at small $T_s$ and $\alpha \simeq 0.5$ at large $T_s$. This crossover has indeed been observed in the experimental walks of Fig.~\ref{fig:expTrack}. The dynamics of the NG can therefore be well-described by our model for small as well as for large time intervals $T_s$. %Further, the observations in Fig.~\ref{fig:hurstS} suggest that the walks for different values of $l_p$ are self-similar.
It may be mentioned that similar exponents can be also be obtained from the root mean squared displacement. The fractal dimension or the Hurst exponent however is defined in the context of the Brownian trace \cite{PA}. 

\begin{figure}[!htbp]
\begin{subfigure}{0.5\linewidth}
\centering
 \includegraphics[scale = 0.29]{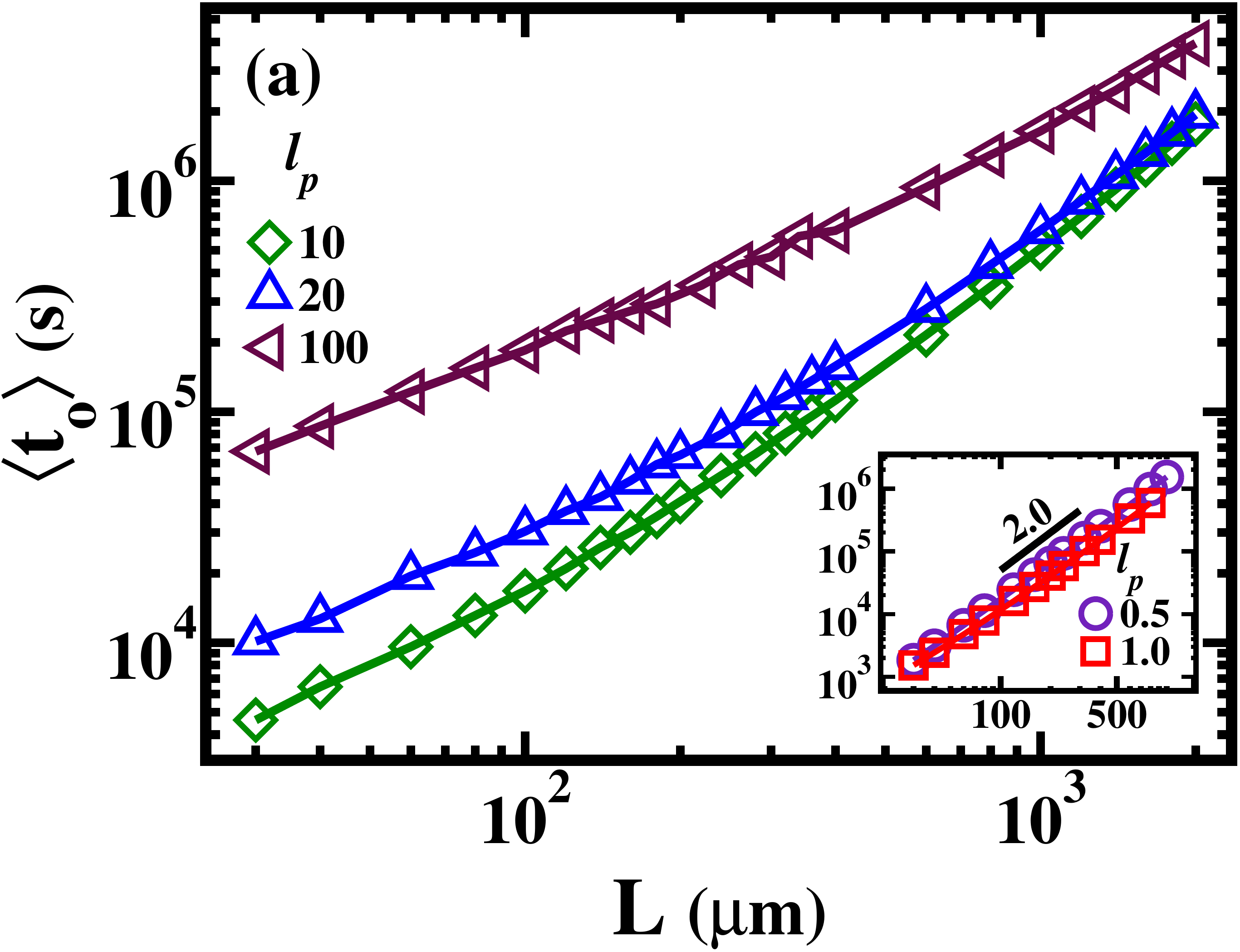}
 \end{subfigure}
\begin{subfigure}{0.49\linewidth}
\centering
\includegraphics[scale = 0.29]{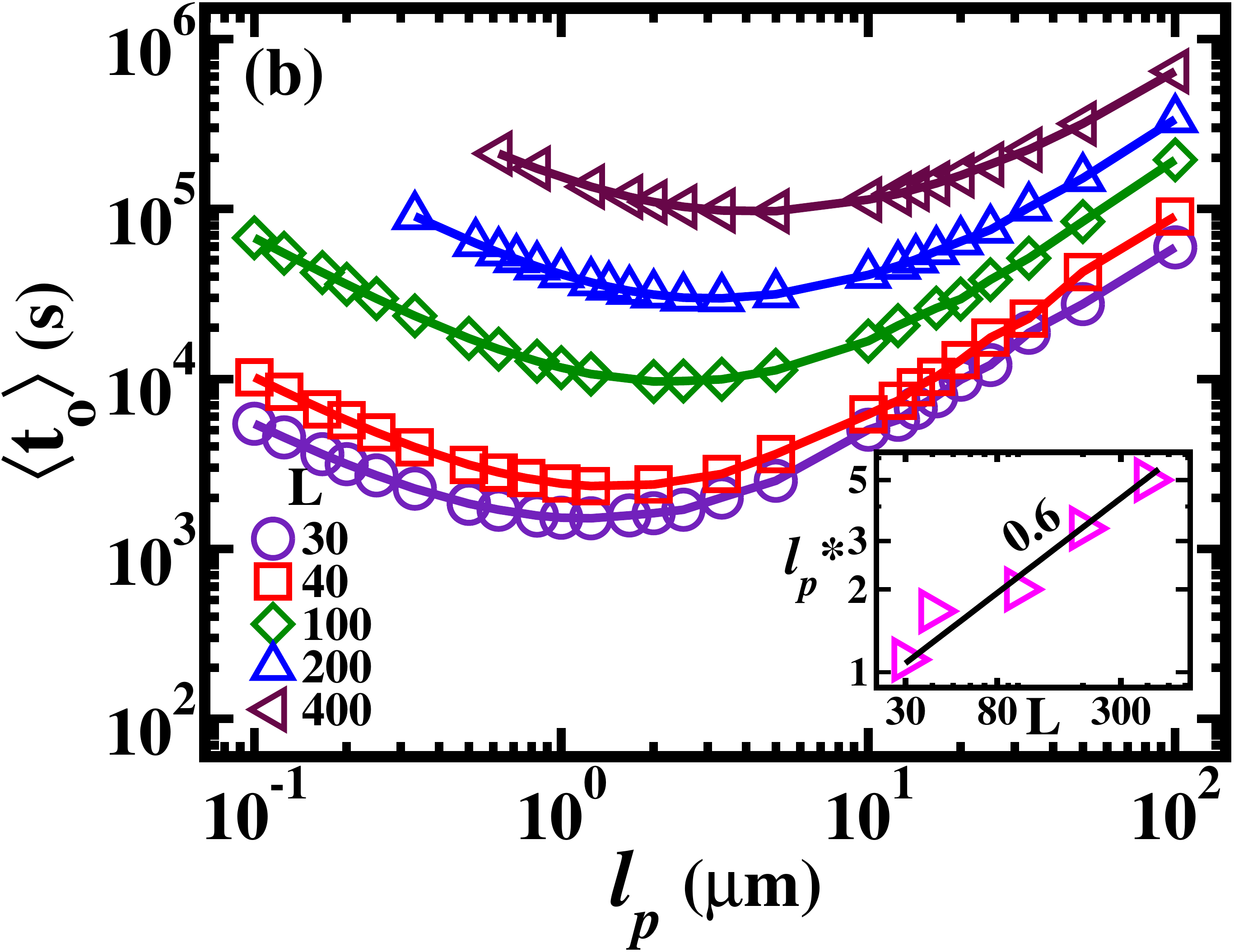}
\end{subfigure}
\caption{(a) FPT $\langle t_o\rangle$ as a function of system size $L$ for $l_p$ ($\mu$m) $= 10, 20, 100$. The inset depicts data for $l_p$ ($\mu$m) $= 0.5$ and $1.0$. The solid line implies that $\langle t_o\rangle\sim L^2$. (b) FPT $\langle t_o\rangle$ as a function of persistence length $l_p$. The inset indicates that the minimum value $l_{p}^{\ast}\sim L^{0.6}$.}
\label{fig:tVs}
\end{figure}

Next we study the dependence of the FPT on $l_p$ and $L$. In our simulations, we always place the target of unit radius at the centre of the $L\times L$ surface, and the walker always starts at a fixed initial position $\left(x_o,y_o\right) = (15~\mu m, 15~\mu m)$. The FPT or $t_o$ is obtained by adding the time taken for each step ($=1/v$) of the random walk which terminates at the target. Note that while calculating the FPT, a step overshooting the target is terminated at it. Further, choosing the same starting point always allows us to study the effect of $L$ on $\langle t_o \rangle$. We also note that this choice is generic, such that the results do not depend strongly on the particular choice. Fig.~\ref{fig:tVs}(a) indicates the variation of $\langle t_o\rangle$ vs. $L$ for $l_p$ = 0.5, 1, 10, 20 and 100 $\mu$m. The angular brackets indicate an averaging over number of walks. For small values of $l_p$ ($\lesssim 1~\mu m$), $\langle t_o\rangle\sim L^2$ characteristic of normal diffusion. This is expected because many small steps, which occur in random directions, are needed to cross a distance of size $L$, over which the trajectories exhibit regular diffusion. Larger values of $l_p$ on the other hand introduce a cross-over towards ballistic motion because distances of order $L$ can be crossed in a directed fashion. Fig.~\ref{fig:tVs}(b) shows the variation of $\langle t_o\rangle $ vs. $l_p$ for $L = $ 30, 40, 100, 200 and 400 $\mu$m. Interestingly, for all values of $L$ there exists an $l_p^*$ which yields the smallest FPT.  As $l_p$ increases from 0, the step lengths become longer due to which walker can reach the target faster till up to $l_p^*$. A further increase however results in the inefficient walk which increases the FPT. The inset depicts the variation of $l_p^*$ vs. $L$. It can be seen that $l_p^*\sim L^{0.6}$.

\subsection{Bacterial Aggregates:}
Finally, we study the formation of bacterial aggregates. Biofilms grow by both simple cell division, and by new free floating microorganisms \cite{ZL, ME, KN}. The resulting morphologies are distinct in the two cases. We believe that such observations are correlated with the distinct walk characteristics on short and long time scales. However, in this work we only study the bacterial aggregates formation and do not take into consideration the bacterial cell division. Aggregation is however an initial stage of more complex biofilm formation process. Their understanding can provide optimized protocols for treatment of infections and also shed light on communication and survival strategies. Experimental observations show that bacterial movement depends on various environmental conditions such as the amount of food (peptone level) or the hardness of the substrate (agar concentration) \cite{DS, JW, EB, FH, MO}. In presence of abundant nutrients and a soft substrate, they move inside the agar and reproduce at a maximal rate. At high agar concentrations, the substrate becomes hard and the bacteria move only on the surface. It has been observed that with poor nutrients and a hard substrate, the colony has a branched diffusion-limited-aggregation (DLA) structure. Adding nutrients yields thicker branches, while soft substrates and high level of nutrients yield a fully formed disc shaped colony. Coarse grained models of bacterial NG colonies have been studied earlier \cite{ZaburdaevPlos15, ZaburdaevPRE15,ZaburdaevNJP17}, but these
studies mainly involved the effect of pili on formation of bacterial colonies. 

\begin{figure}[!htbp]
\begin{subfigure}{.34\linewidth}
 \centering
 \includegraphics[scale = 0.25]{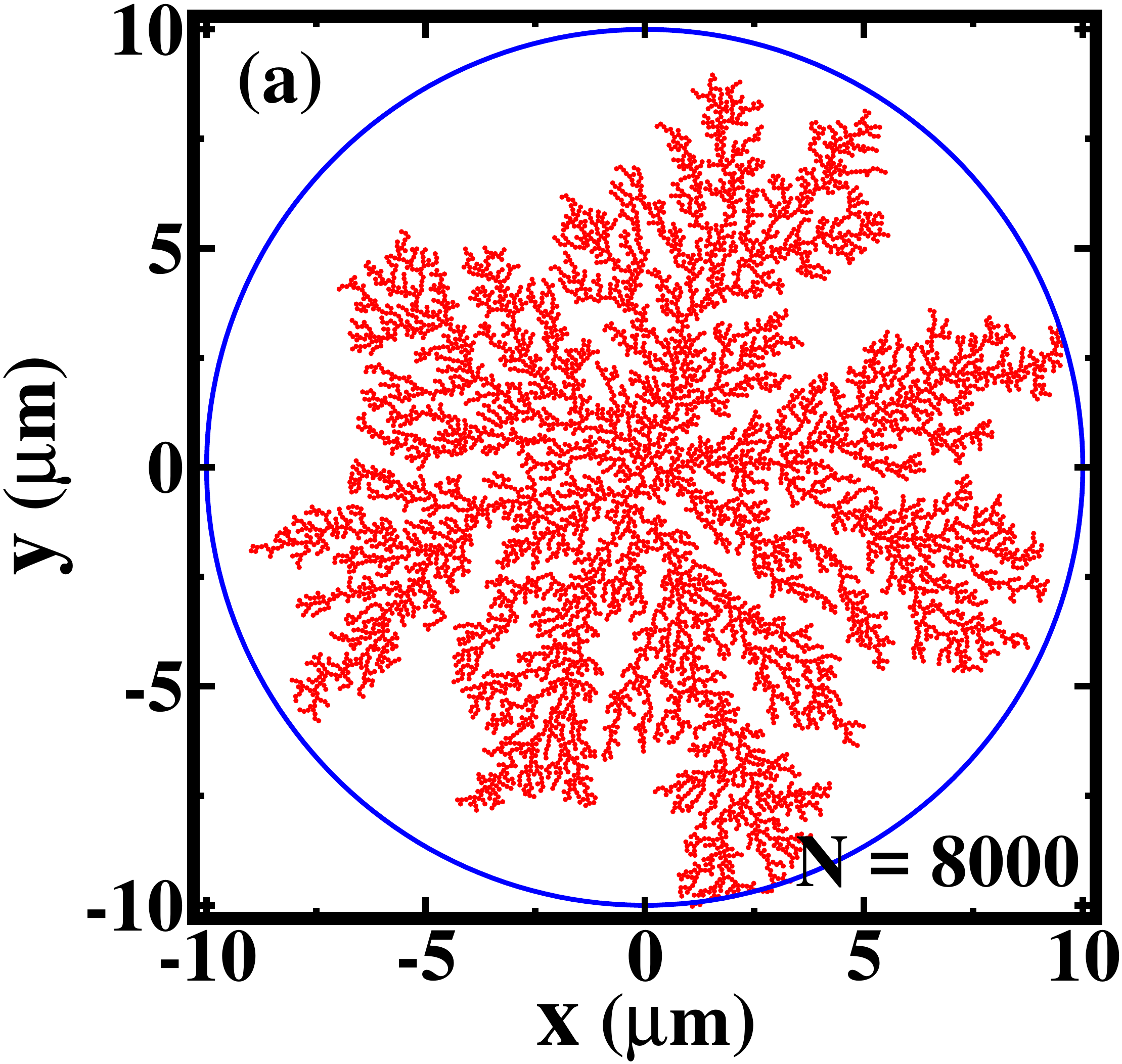}
 \end{subfigure}
\begin{subfigure}{.32\linewidth}
\centering
\includegraphics[scale = 0.25]{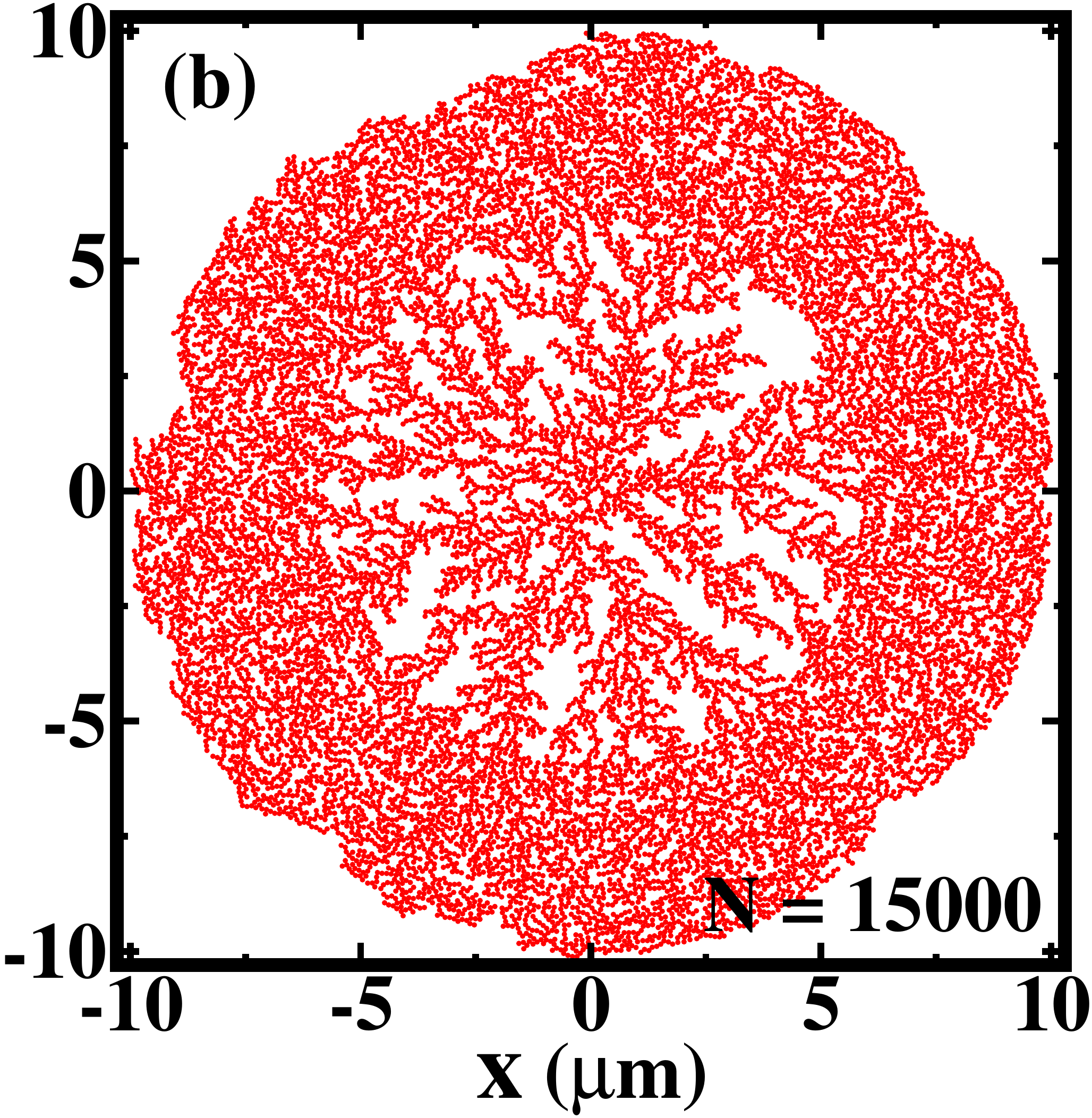}
\end{subfigure}
\begin{subfigure}{.32\linewidth}
\centering
\includegraphics[scale = 0.25]{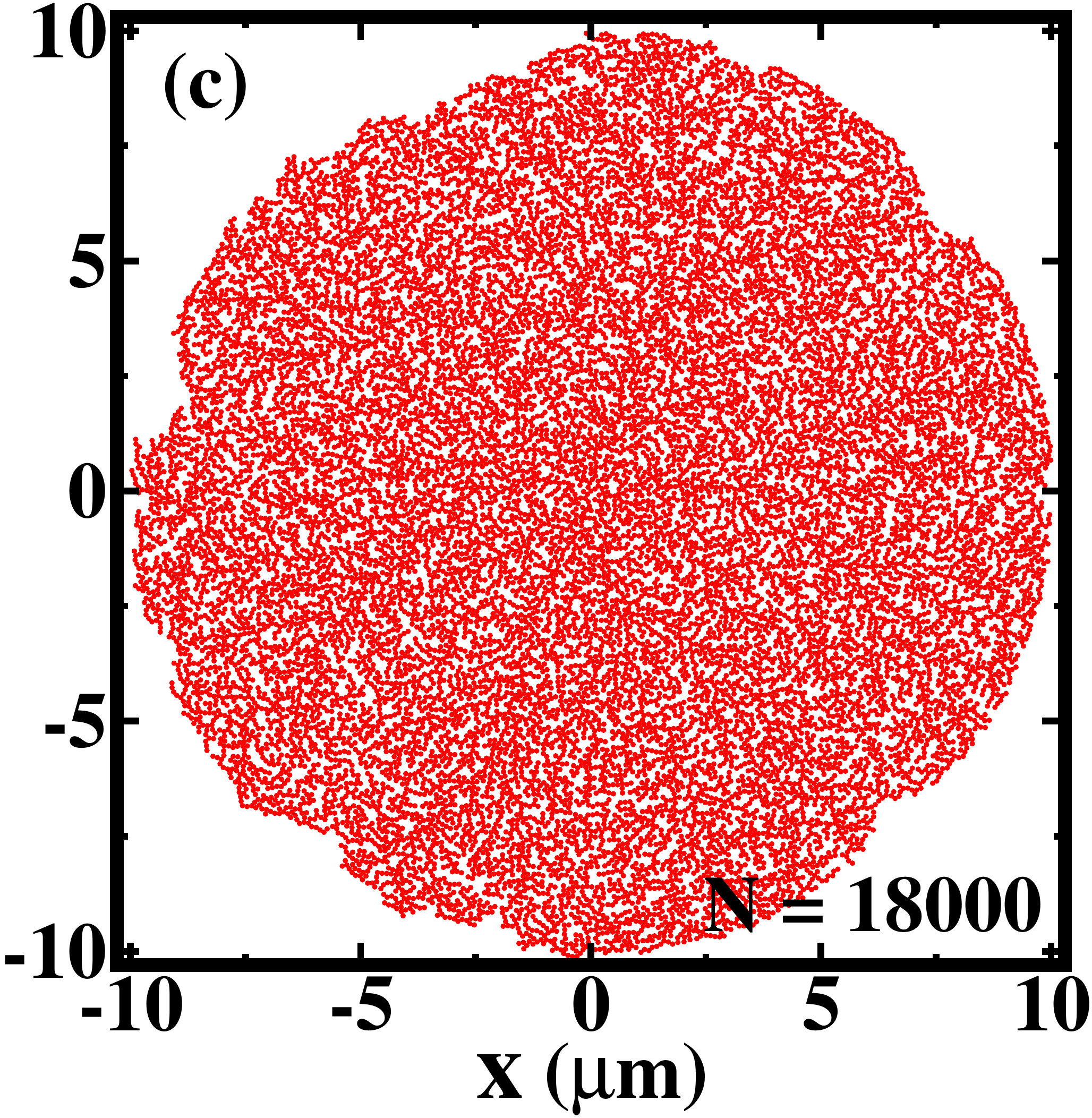}
\end{subfigure}
\begin{subfigure}{0.33\linewidth}
  \vspace{0.5cm}
 \centering
 \includegraphics[scale = 0.2]{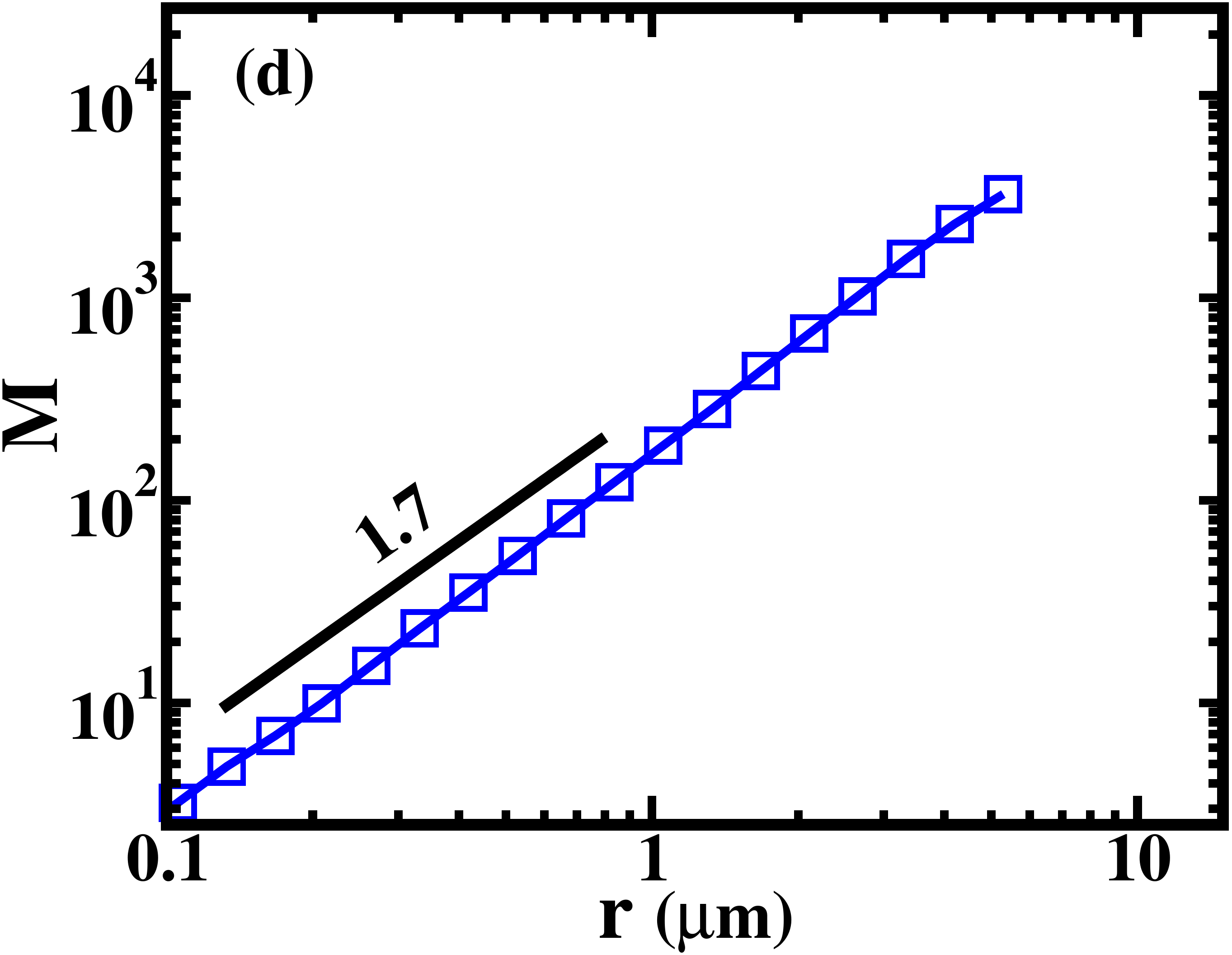}
 \end{subfigure}
 \begin{subfigure}{0.32\linewidth}
   \vspace{0.5cm}
\centering
\includegraphics[scale = 0.2]{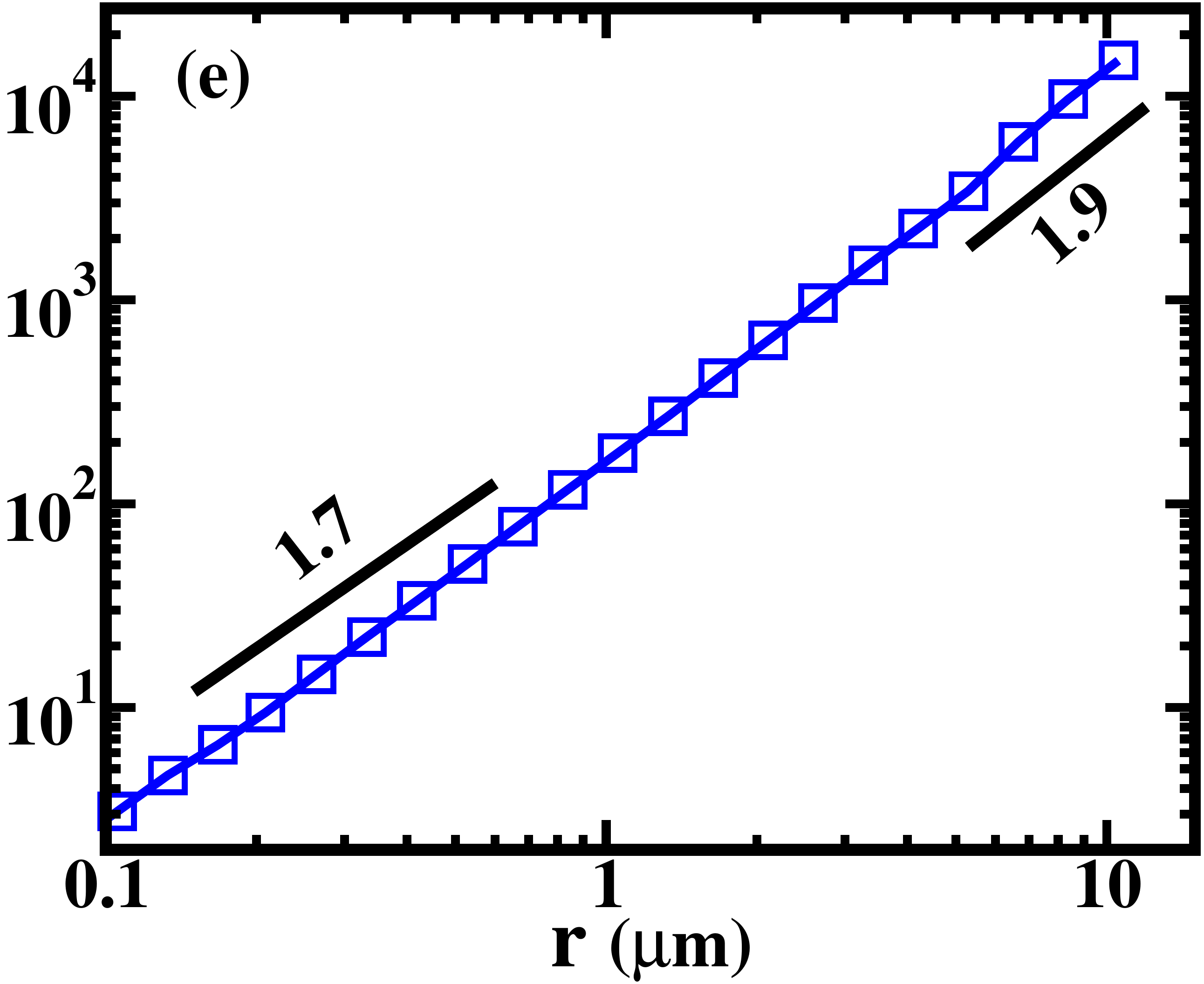}
\end{subfigure}
\begin{subfigure}{0.32\linewidth}
  \vspace{0.5cm}
\centering
\includegraphics[scale = 0.2]{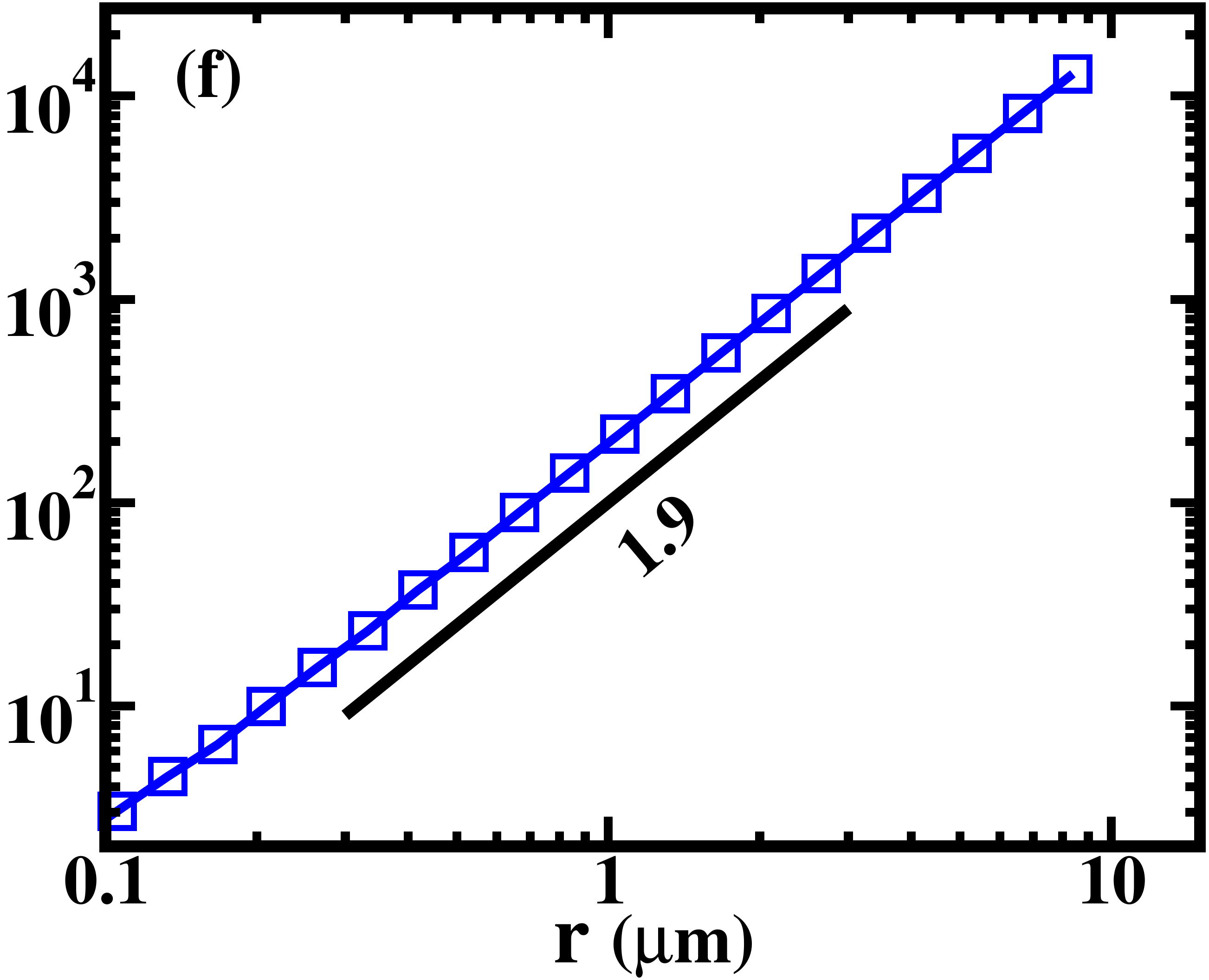}
\end{subfigure}
\caption{Evolution of bacterial aggregates (refer text) for $l_p = 1$ $\mu$m: (a) DLA structure at early time with $N = 8000$ bacteria, (b) at intermediate time with $N = 15000$ bacteria and (c) at late time with $N = 18000$ bacteria. Notice the evolution from a delicate DLA morphology with fractal structure to a dense morphology with little structure. The lower panel (d), (e) and (f) depicts corresponding fractal dimension of morphologies in (a), (b) and (c) respectively.}
\label{fig:dla}
\end{figure} 

However, our interest here is to observe the effect of the short time directed ballistic motion $\left(\alpha \simeq 1\right)$ and the long time regular diffusion $\left(\alpha \simeq 0.5\right)$ on the cluster morphology. We use a simple procedure to create the aggregates. In our simulations, bacteria are treated as circular particles of diameter $0.1$ $\mu$m. The initial position of the bacterium is randomly chosen on a circle of radius $R_d = 10$ $\mu$m with the nucleating site at the center. The bacterium executes a random walk with steps drawn from the exponential distribution of Eq.~(\ref{exp}). On reaching the nucleating site, it adheres irreversibly. A new walk is initiated and the process continues to yield a DLA cluster. For computational efficiency, we only consider those walks which are within the circle. The evolution of the cluster is depicted in Fig.~\ref{fig:dla}(a)-(c). In (a), notice that the characteristic DLA backbone is created first. It results from the long random walks extending from the circumference to the centre. The time taken for these walks $\gtrsim 10^4$ s, which is typically the time interval over which $\alpha \simeq 0.5$, see Fig.~\ref{fig:hurstS}. As the dendrites grow, (b) indicates a distinct change in the morphology at the periphery. The irreversible aggregation happens over $<10^3$ s which corresponds to the fast ballistic diffusion regime with $\alpha \simeq 1$. We therefore conclude that while the DLA backbone is formed by the slowly diffusing bacteria, the thickening is due to the fast ballistic or directed motion at shorter time scales.

In order to characterize these aggregates, we calculate their fractal dimension $d_f$ \cite{JCR} using the relation: 
\begin{equation}
M(r)\sim r^{d_f}.
\label{df}
\end{equation} 
The mass $M(r)$ is calculated within a circle of radius $r$ by incrementing the mass by unity for each particle. To facilitate evaluation of  the fractal dimension of the dense morphology accurately, we create the filled structure in (c) by systematically reducing the drop radius. The corresponding evaluations of the fractal dimension for morphologies in Fig.~\ref{fig:dla}(a)-(c), are depicted in lower panel of Fig.~\ref{fig:dla}: (d) $d_f \simeq 1.7$ for the DLA-like structure and (f) $d_f \simeq 1.9$ for the dense morphology. The intermediate time morphology in Fig.~\ref{fig:dla}(e) depicts a crossover between the two regimes due to the vastly different morphologies in the bulk and the periphery.  

\begin{figure}[!htbp]
\includegraphics[scale=0.27]{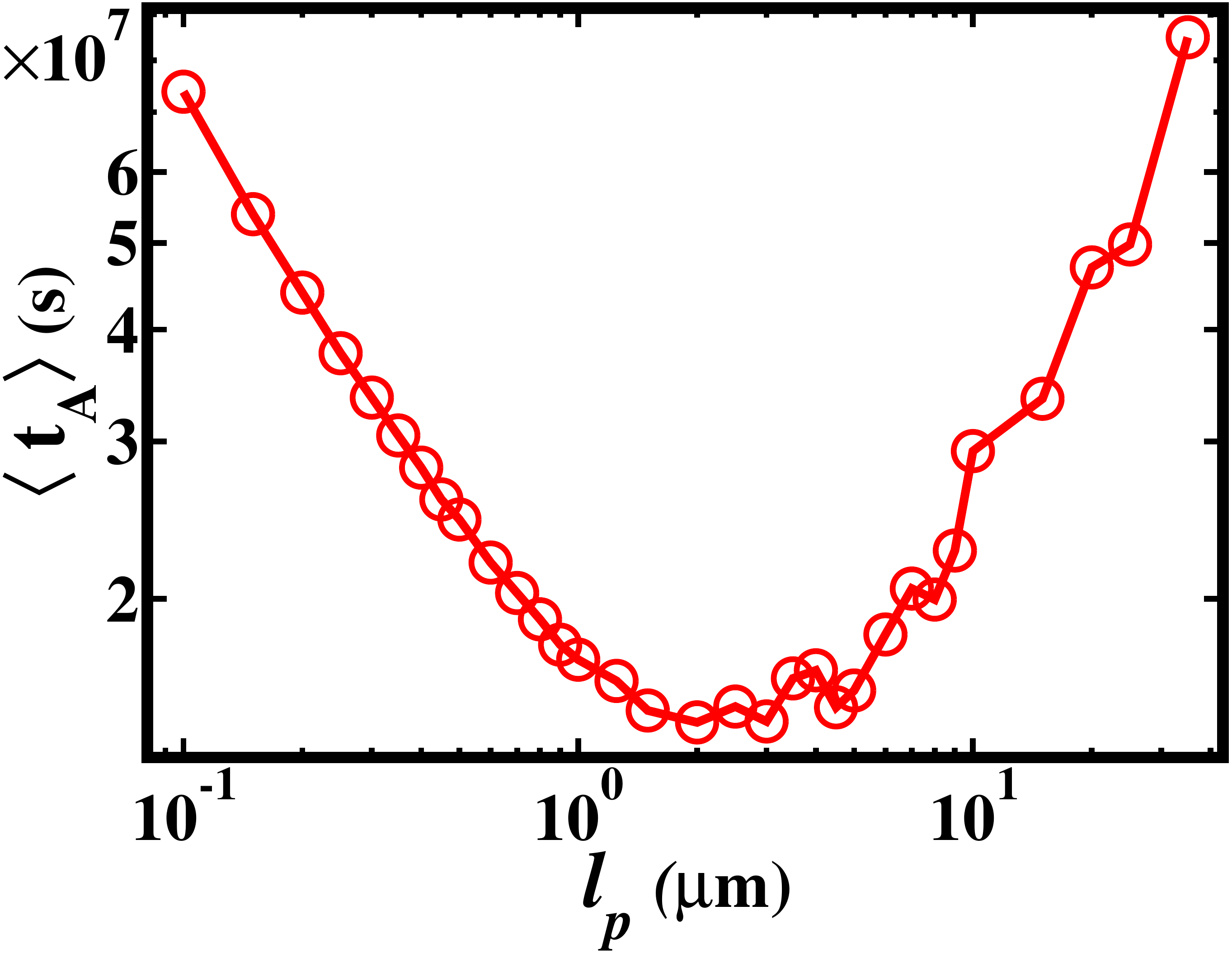}
\caption{Average total time $\langle t_A\rangle$ taken for the formation of bacterial colony of 
$N=5000$ bacteria on a $L\times L$ with $L=100~\mu m$ surface, drop radius $R_d=10~\mu m$, 
with along the wall boundary condition at the boundaries of the surface, discussed in the text. 
The average is done over $30$ configurations. A distinct minimum can be seen at $l_p\sim 1~\mu m$, 
as in case of the FPT (see Fig. \ref{fig:tVs}(b)).}
\label{fig:tADLA}
\end{figure}

We expect there exists a characteristic length $l_p^*$, which leads to the fastest growth of the aggregates. We thus simulated microcolonies with along the wall boundary conditions at the boundaries of a $L\times L$ surface with $L=100 \mu m$, and drop radius $R_d=10~\mu m$. We then measured the average total time $\langle t_A\rangle $ taken for the formation of bacterial colonies of $N=5000$ bacteria. This is a typical number for which aggregates remain well within the drop radius. Results are shown in Fig. \ref{fig:tADLA}, where a distinct minimum at $l_p\sim 1~\mu m$ is seen which matches with
the minimum seen in FPT.

\begin{figure}[!htbp]
\vspace{0.5cm}
\includegraphics[scale = 0.29]{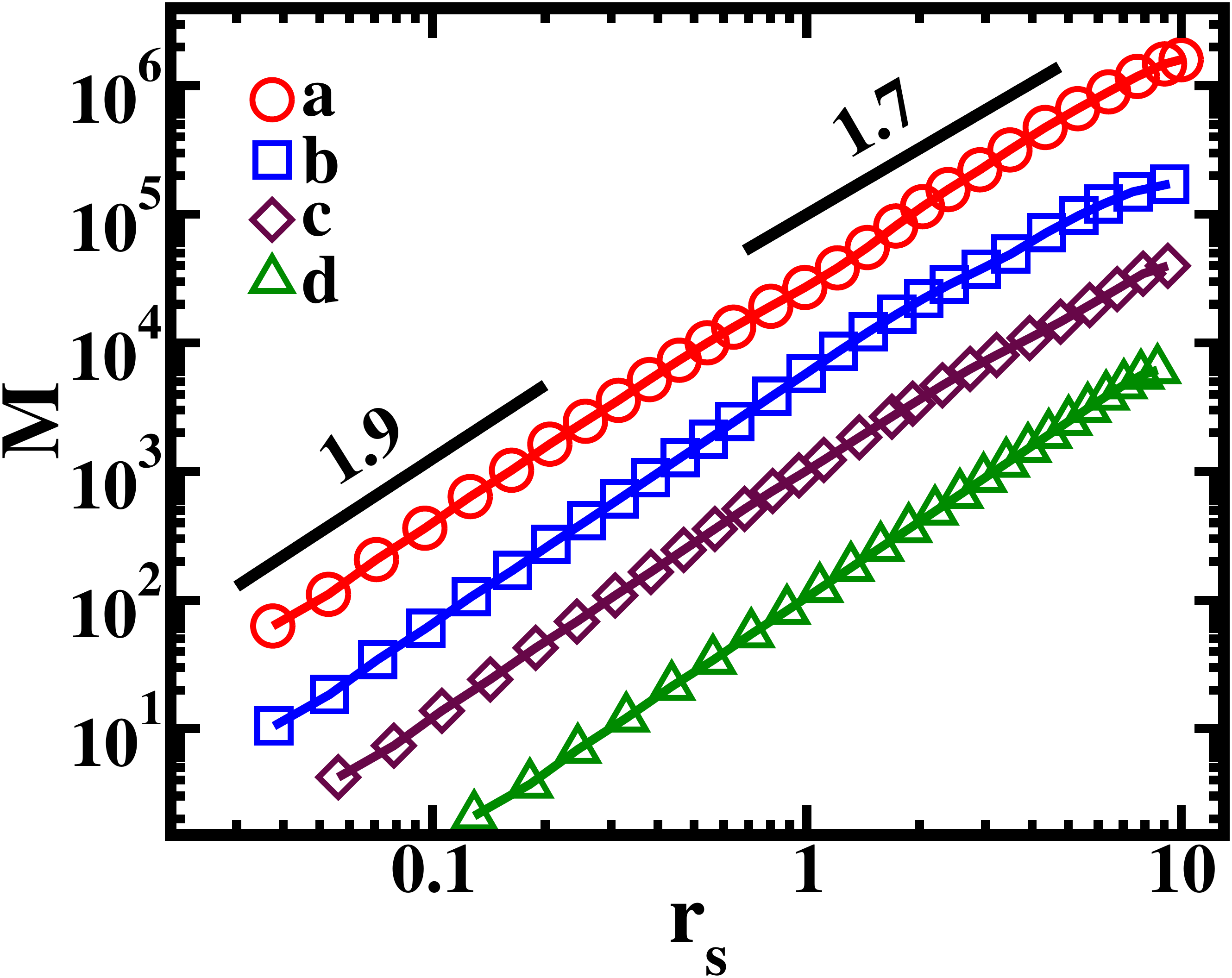}
\caption{Evaluation of fractal dimension of several bacterial colonies at different environmental conditions of (a) \textit{Paenibacillus vortex} on $2.25$\% agar with $2$\% peptone \cite{CJI}, (b) \textit{Bacillus subtilis} with $1$ gl$^{-1}$ peptone \cite{MM}, (c) \textit{Paenibacillus dendritiformis} on $1.75$\% agar with $0.5$ gl$^{-1}$ peptone \cite{EBJ}, and (d) \textit{Serratia marcescens} on $1.4$\% agar with $0.4$\% glucose \cite{TM}. The mass $M$ within a circle of radius $r$ as a function of rescaled radial distance $r_s$ where $r_s = 10\times r/r_{o}$ and $r_{o}$ is the radius of the aggregate. Then using Eq.~(\ref{df}), we obtain $d_f$ for these aggregates. The data sets have been shifted for clarity. The solid lines with slopes $1.7$ and $1.9$ are a guide to the eye.}
\label{fig:mVsRE}
\end{figure}

We have also analyzed the experimentally determined morphologies of microcolonies \cite{CJI, MM, EBJ, TM}, for a comparison with the results of our model. However, in these experiments, the bacteria were injected at the center of the petri dish and allowed to diffuse. Therefore, the microcolonies are denser near the center but fractal at the periphery due to bacterial diffusion. Thus the morphology is complementary to that generated in our simulations where injection is at the periphery. In Fig.~\ref{fig:mVsRE}, we evaluate the fractal dimension of the microcolonies obtained under different environmental conditions from their snapshots (see caption for details). Thus, we find, $d_f\simeq 1.9$ for the morphology in the center, while $d_f\simeq 1.7$ for the delicate branched structures.

\section{Summary and Discussion}
In this paper, we have analyzed the twitching motility of bacteria with Type IV pili such as the Neisseria gonorrhoeae (NG). The step lengths in the NG trajectories obey an exponential distribution: $P(l) = (1/l_p) \exp\left(-l/l_p\right)$ with the persistence length $l_p\simeq 2$ $\mu$m. The Hurst exponent evaluated from these tracks exhibits a crossover from a ballistic regime ($\alpha \simeq 1.0$) to a diffusive regime ($\alpha \simeq 0.5$) as the window of observation increases. The walks are {\it fractal} and {\it persistent}. The bacterial motion is a superposition of fast ballistic diffusion (directed motion) on small time scales and slow diffusion (random walk) over longer time scales. Computer simulations reveal that for each value of $L$, there is an optimal value of the persistence length $l_p^*$ for which the first passage time (FPT) is the least. In experiments, NG bacterium  walks can be tuned to have a range of values of $l_p$. So, it will be interesting to see if indeed NG walks show a minimum in $\langle t_o\rangle$ for some value of $l_p$. The optimal value $l_p^*$, 
also leads to the fastest growth of the bacterial aggregates. Such a minimum could have significance with respect to bacterial fitness and could thus be evolvable. Many bacteria search for a target such as a food source or an emergent colony of their species using chemotaxis, sensing a chemical gradient. However, not all bacteria show chemotaxis. NG is an example that lacks chemotaxis genes. Unless they are guided by mechanical properties of the surface, these bacteria have to find the target by an unbiased random walk. The existence of an optimal persistence length suggests that the search in these bacteria could be optimized by an evolutionary mechanism if in their natural environment they face a characteristic confinement length.

We have also studied the formation of bacterial aggregates or biofilms. These resemble a diffusion-limited-aggregation (DLA) cluster in the early stages. Morphology of the cluster changes drastically as it grows and one requires different fractal dimensions to characterize the core and the periphery. This complements the crossover seen in the Hurst exponent. Morphologies indicate that the DLA backbone is a result of diffusive motion at longer time scales and dense structures are due to ballistic motion. We also observe similar behavior in a few experimental bacterial colonies.

\acknowledgments
 Authors also thank B. Maier and C. Meel for providing the experimental trajectories depicted in Fig.~\ref{fig:expTrack}(a). KB acknowledges CSIR India for financial support under the grant number 09/086(1208)/2015-EMR-I. VB acknowledges DST India for a research grant. RM acknowledges DST-INSPIRE Faculty scheme for a research grant. The authors gratefully acknowledge the HPC facility of IIT Delhi for
computational resources.

\end{document}